\documentclass[prl,twocolumn,showpacs,amsmath,amssymb,superscriptaddress, nofootinbib]{revtex4-1}
\usepackage{epsfig,epstopdf} % Transform eps to pdf
\usepackage{graphicx}% Include figure files
\usepackage{dcolumn}% Align table columns on decimal point
\usepackage{bm}
\usepackage{bbm}
\usepackage{ulem}
\usepackage{xcolor}
\usepackage{amsmath}
\usepackage{esint}
\usepackage{bbold}
\usepackage{slashed}
\usepackage{mathrsfs}
\usepackage{subfigure}
\usepackage{verbatim}
\usepackage{dsfont}
\usepackage{float}
\usepackage{wasysym}
\usepackage[utf8]{inputenc}

\usepackage[breaklinks]{hyperref}
\hypersetup{colorlinks=true, linkcolor=blue, citecolor=blue, filecolor=blue, urlcolor=blue}

\AtBeginDocument{%
	\newwrite\bibnotes
	\def\bibnotesext{Notes.bib}
	\immediate\openout\bibnotes=\jobname\bibnotesext
	\immediate\write\bibnotes{@CONTROL{REVTEX41Control}}
	\immediate\write\bibnotes{@CONTROL{%
			apsrev41Control,author="08",editor="1",pages="1",title="0",year="1"}}
	\if@filesw
	\immediate\write\@auxout{\string\citation{apsrev41Control}}%
	\fi
}%

\begin{document}

\title{Anomalous Floquet Chiral Topological Superconductivity in a Topological Insulator Sandwich Structure}
\author{Rui-Xing Zhang}
\email{ruixing@umd.edu}
\author{S. Das Sarma}
\affiliation{Condensed Matter Theory Center and Joint Quantum Institute, Department of Physics, University of Maryland, College Park, Maryland 20742-4111, USA}

\begin{abstract}
	We show that Floquet chiral topological superconductivity arises naturally in Josephson junctions made of magnetic topological insulator-superconductor sandwich structures. The Josephson phase modulation associated with an applied bias voltage across the junction drives the system into the anomalous Floquet chiral topological superconductor hosting chiral Majorana edge modes in the quasienergy spectrum, with the bulk Floquet bands carrying zero Chern numbers. The bias voltage acts as a tuning parameter enabling novel dynamical topological quantum phase transitions driving the system into a myriad of exotic Majorana-carrying Floquet topological superconducting phases. Our theory establishes a new paradigm for realizing Floquet chiral topological superconductivity in solid-state systems, which should be experimentally directly accessible. 
\end{abstract}
\date{\today}
\maketitle

{\it Introduction} - Ever since the discovery of the quantum Hall effect \cite{klitzing1980,thouless1982}, the concept of topological phases has revolutionized our understanding of matters by challenging the conventional Landau paradigm of classifying phases of matter. Two phases, sharing exactly the same internal and crystalline symmetries, could still be distinct in a topological sense, thus behaving differently in terms of their physical properties. Besides the quantum Hall effect, other well-known examples of topological phases include Chern insulators \cite{haldane1988model,liu2016QAH}, topological insulators \cite{kane2005z2,bernevig2006QSH,hasan2010colloquium,qi2011topological} and topological superconductors (TSCs) \cite{read2000paired,Kitaev_2001,lutchyn2010majorana}. In particular, the intrinsic connection between TSCs and non-Abelian Majorana zero modes has inspired tremendous research activity spanning condensed matter physics to quantum computation \cite{sarma2015majorana,Lutchyn2018,Prada2020}. The current work introduces a new idea involving driven dynamical Floquet chiral Majorana modes in TSCs.

Recently, far-from-equilibrium dynamical topological phenomena have attracted research attention \cite{oka2009photovoltaic,lindner2011floquet,cayssol2013floquet,oka2019floquet,harper2020topology,zhang2020tunable}. In particular, Floquet systems subjected to time-periodic dynamical driving may manifest new topological phases emergent as a consequence of the quantum dynamics. These Floquet systems may host boundary modes which are simply impossible in static equilibrium systems \cite{kitagawa2010topo,jiang2011majorana,rudner2013anomalous,nathan2015topo,morimoto2017floquet,yao2017topo,peng2019floquet,zhang2020theory}. A well-known ``beyond-static-topology" example is a two-dimensional (2D) anomalous Floquet topological insulator (AFTI) in the symmetry class A \cite{kitagawa2010topo,rudner2013anomalous}. Remarkably, the boundary of such an AFTI hosts chiral topological modes just like a static Chern insulator, even though all its Floquet bands are topologically trivial. Such an AFTI phase has already been experimentally demonstrated in the state-of-the-arts photonic  \cite{mukherjee2017experimental,maczewsky2017observation,adzal2020realization,mukherjee2020observation}, acoustic \cite{peng2016experimental}, and atomic \cite{Wintersperger2020} platforms. Despite the success in realizing AFTIs, however, there have been surprisingly few theoretical studies \cite{yang2018floquet} and no experimental efforts on the corresponding superconducting counterpart, i.e. a 2D anomalous Floquet chiral topological superconductor (AFCTSC) with Chern-number-independent chiral Majorana edge modes. Comparing to AFTIs, the key challenge to materialize an AFCTSC phase is two-fold: (i) the intrinsic difficulty in simulating superconductivity in artificial systems; (ii) the lack of a simple and feasible dynamical driving protocol in superconducting solid-state systems.

\begin{figure}[t]
	\includegraphics[width=0.49\textwidth]{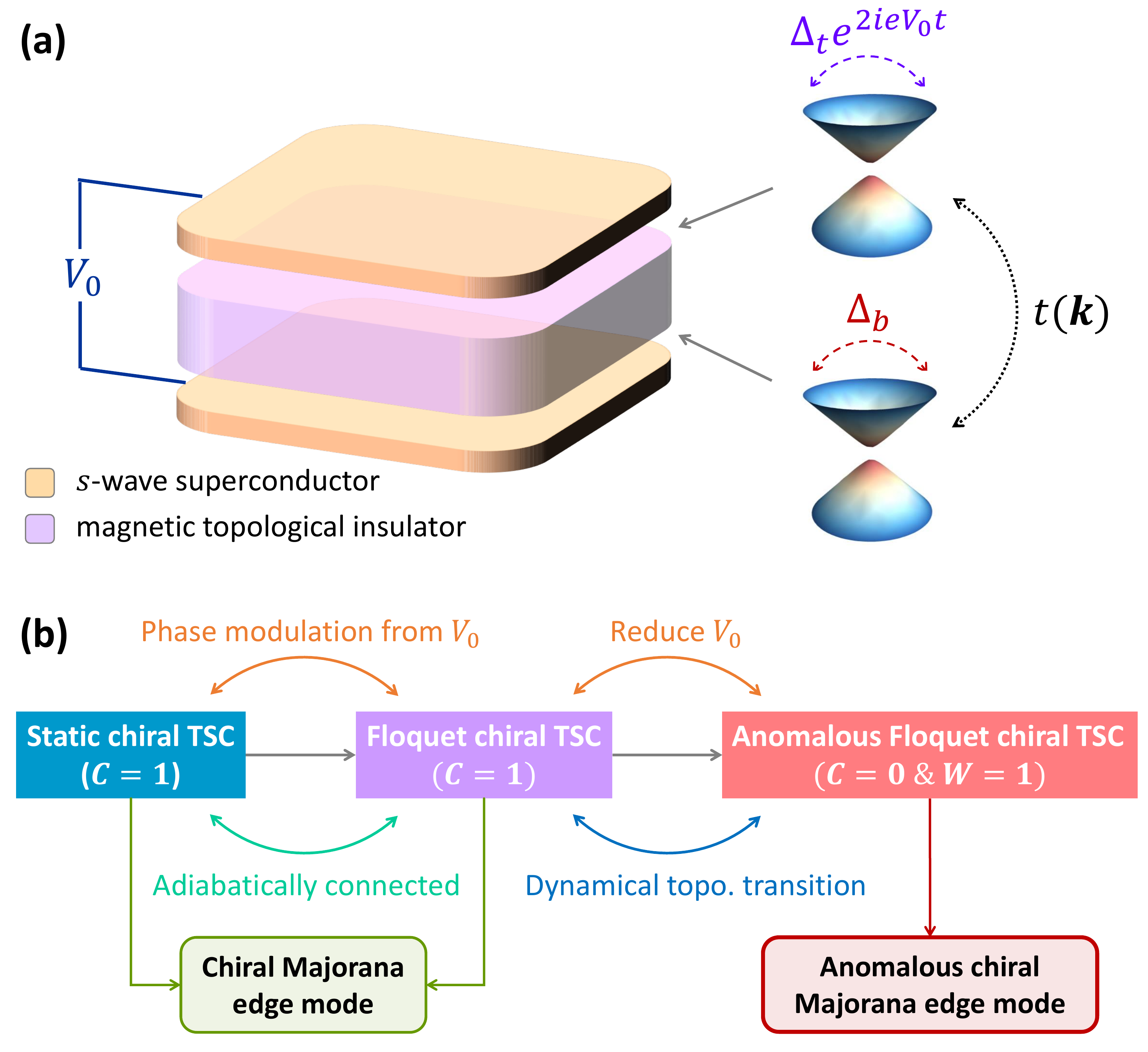}
	\caption{(a) A chiral topological Josephson junction. The low-energy physics is governed by the gapped Dirac surface states of the magnetic topological insulator thin film, along with the intersurface tunneling $t({\bf k})$ and the proximitized superconducting orders $\Delta_{t, b}$. Applying a bias voltage $V_0$ achieves a modulation of the Josephson phase $\varphi(t)$ and drives the system into a Floquet superconductor. (b) Tuning $V_0$ induces dynamical topological phase transition, which further leads to anomalous Floquet chiral TSC phase. The anomalous chiral Majorana edge modes denote those that cannot be explained by the BdG Chern number of the Floquet bulk bands.}
	\label{Fig1}
\end{figure} 

In this work, we propose an experimentally feasible and highly tunable solid-state paradigm for achieving AFCTSC and Floquet chiral Majorana physics. Our proposed platform is based on a superconductor/magnetic topological insulator/superconductor sandwich structure [see Fig. \ref{Fig1} (a)], which we call a {\it chiral topological Josephson junction} (CTJJ). With a static bias voltage applied across the top and bottom superconducting layers, the Josephson phase (i.e. the relative phase between the superconductors) modulates as a periodic function of time, offering a driving protocol for this CTJJ. Remarkably, by simply tuning the DC bias voltage, a series of dynamical topological phase transitions occur sequentially, as schematically shown in Fig. \ref{Fig1} (b). This leads to a wealth of novel Floquet TSC states, including the AFCTSC state with anomalous chiral Majorana edge modes. We provide a detailed theory for the emergent Floquet topological phases, including their dynamical formation and topological characterization.

{\it Model Hamiltonian} - We start by introducing the model Hamiltonian for the CTJJ system (Fig. \ref{Fig1}). A key ingredient here is the magnetic topological insulator thin film, which can be effectively described by $h_\text{MTI} ({\bf k}) = v(\sin k_x \sigma_z \otimes s_y - \sin k_y \sigma_z \otimes s_x) + t({\bf k}) \sigma_x + g_z s_z$. Here $\sigma$ and $s$ are the Pauli matrices denoting the surface layer and spin degrees of freedom, respectively. $g_z$ denotes the $z$-directional Zeeman effect due to the bulk magnetization and $t({\bf k}) = t_0 - t_1 (\cos k_x + \cos k_y)$ describes the hybridization between top and bottom Dirac surface states. In the presence of the superconductor layers, the effective Bogoliubov-de Gennes (BdG) theory for the system is given by
\begin{eqnarray}
H_\text{CTJJ}({\bf k}, t) &=& \begin{pmatrix}
h_\text{MTI}({\bf k}) -\mu & h_\Delta (t) \\
h_\Delta^{\dagger} (t) & -h_\text{MTI}^T(-{\bf k}) +\mu
\end{pmatrix} \nonumber \\
\end{eqnarray} 
where the isotropic $s$-wave pairing is given by
\begin{equation}
	h_\Delta (t) = \begin{pmatrix}
	-i \Delta_t e^{i\varphi(t)} s_y & 0 \\
	0 & -i \Delta_b s_y
	\end{pmatrix}.
\end{equation}
Here, $\Delta_{t (b)}$ is the proximitized superconducting orders for the top (bottom) surfaces of the magnetic topological insulator and $\varphi(t)$ is the Josephson phase between $\Delta_t$ and $\Delta_b$. $\tau_{0,x,y,z}$ are the Pauli matrices for the particle-hole degree of freedom and we define $\sigma_{\pm} = (\sigma_0 \pm \sigma_z)/2$ and $\mu$ as the chemical potential.

When a constant bias voltage $V_0$ is applied between the superconductor layers, $\varphi(t)$ starts to periodically modulate according to the second Josephson relation
\begin{equation}
	\varphi(t) = \varphi_0+ 2eV_0 t.
\end{equation} 
This converts the CTJJ to an effective Floquet superconducting system with a driving frequency of $\omega = 2eV_0$, although the dynamics here is intrinsically generated by the Josephson effect and not by any explicit external time dependent field. In the frequency domain, the Floquet Hamiltonian ${\cal H}_F$ is an infinite-dimensional matrix 
%\begin{equation}
%	{\cal H}_F({\bf k}) = \begin{pmatrix}
%	\ddots & \vdots & \vdots & \iddots \\
%	\dots & h_0 -\omega & h_{1} & \dots \\
%	\dots & h_{-1} & h_0 & \dots  \\
%	\iddots & \vdots & \vdots & \ddots  \\
%	\end{pmatrix}
%\end{equation}
with $({\cal H}_F)_{ll'} = h_\omega^{(l-l')} + \omega l\delta_{ll'}$ for $l,l'\in\mathbb{Z}$. Here we have defined $h_\omega^{(l-l')}=1/T\int_0^T H_\text{CTJJ}(t)e^{i(l-l')\omega t} dt$ with the driving period $T=2\pi/\omega$. It is straightforward to show that
\begin{eqnarray}
	h_\omega^{(0)} &=& H_\text{CTJJ}(\Delta_t \rightarrow 0), \nonumber \\
	h_\omega^{(1)} &=& \frac{\Delta_t}{2} (\tau_y \otimes \sigma_+ \otimes s_y - i\tau_x \otimes \sigma_+ \otimes s_y),
\end{eqnarray} 
along with $h_\omega^{(-1)}=(h_\omega^{(1)})^\dagger$ and $h_\omega^{(l)}=0$ if $|l|>1$.

%To evaluate the quasienergy spectrum of ${\cal H}_F$, we can take a finite truncation of this infinite matrix.

{\it High-frequency Limit} - When the energy scale of $eV_0$ is much larger than the bandwidth of $h_\omega^{(0)}$, the system enters the high-frequency regime and the corresponding topological physics is governed by $h_\omega^{(0)}$, the zeroth order contribution in the Floquet-Magnus expansion of ${\cal H}_F$.

When $\mu=0$, the eigenspectrum of $h_\omega^{(0)}$ is analytically tractable with
\begin{equation}
E^{(0)}_\omega=\pm \sqrt{v^2 k^2 + (g_z \pm \frac{\Delta_b}{2} \pm \frac{\sqrt{\Delta_b^2 + 4 t({\bf k})^2}}{2})},
\end{equation}     
where $k^2=k_x^2+k_y^2$. Then it is straightforward to show that the bulk gap-closing condition for $h_\omega^{(0)}$ at a high-symmetry momentum $k_i$ is simply
\begin{eqnarray}
    \Delta_b = \pm \frac{t(k_i)^2-g_z^2}{g_z}.
\end{eqnarray}
By defining $\alpha=2t_1/t_0$, we have $t(\Gamma) = (1+\alpha)t_0,\ t(X/Y) = t_0,\ t(M) = (1-\alpha)t_0$ for the high-symmetry momenta $\Gamma=(0,0), X=(\pi,0), Y=(0,\pi)$, and $M=(\pi,\pi)$. Therefore, the topological phase boundaries for $h_\omega^{(0)}$ at $\mu=0$ are completely determined by three independent parameters: $g_z,\Delta_b$, and $\alpha$.

\begin{figure}[t]
	\includegraphics[width=0.49\textwidth]{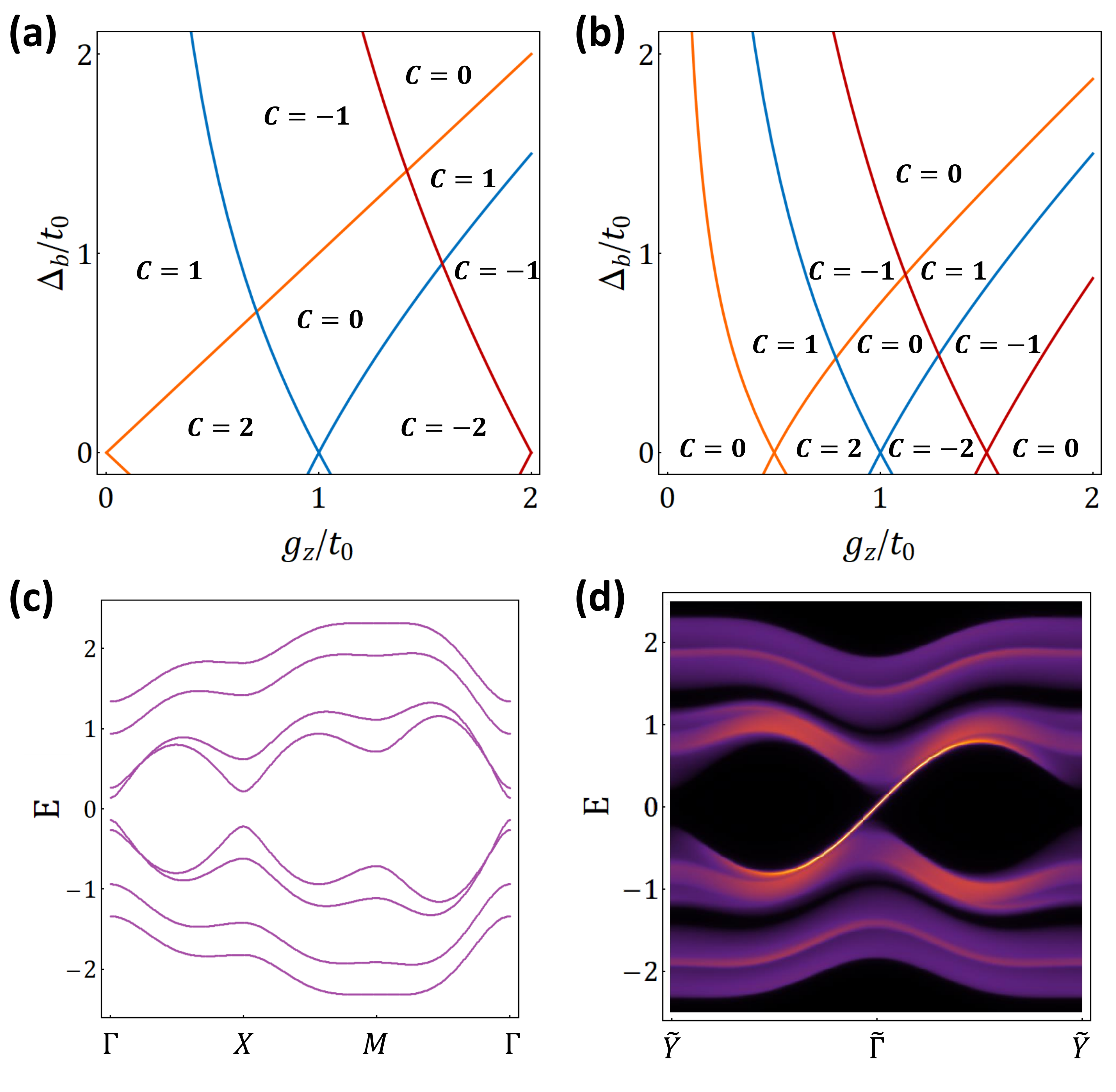}
	\caption{(a) \& (b) show the topological phase diagram of $h_\omega^{(0)}$ for $\alpha=1$ and $\alpha=\frac{1}{2}$, respectively. The orange, blue, and red phase boundaries denote band gap closing at zero energy at $\Gamma$, $X$ \& $Y$, and $M$, respectively. (c) Bulk band spectrum of $h_\omega^{(0)}$ for our choice of parameters with the BdG Chern number $C=1$. (d) Edge spectrum of the $C=1$ phase, which clearly shows a single chiral Majorana edge mode. $\widetilde{\Gamma}$ and $\widetilde{Y}$ are the high-symmetry momenta in the edge BZ.}
	\label{Fig2}
\end{figure} 

In Fig. \ref{Fig2} (a) and (b), we map out the topological phase diagram for $h_\omega^{(0)}$ with $\alpha=1$ and $\alpha=\frac{1}{2}$, respectively. We have identified a plethora of chiral topological superconducting phases with a non-zero BdG Chern number $C$ for the occupied band manifold, all of which feature $|C|$ number of chiral Majorana edge modes. Whenever a single Dirac transition occurs at $\Gamma$ or $M$ (denoted by the orange and red lines, respectively), we find that the BdG Chern number $C$ changes by $\pm1$. Since our model is invariant under the four-fold rotation symmetry, the bulk gap will simultaneously close at $X$ and $Y$ (denoted by the blue line), changing the value of $C$ by $\pm2$. In Fig. \ref{Fig2} (c) and (d), we choose the parameter set $\mu=0, v=0.8, \Delta_b=0.4, g_z=0.6, \alpha=0.5$ in unit of $t_0$ and plot the bulk and edge spectra of $h_\omega^{(0)}$, respectively. These parameters should generate a $C=1$ chiral TSC phase based on Fig. \ref{Fig2} (b), which is further confirmed by the chirally dispersing Majorana edge mode shown in Fig. \ref{Fig2} (d).

Notably, the phase diagram of $h_\omega^{(0)}$ should be directly interpreted as the Floquet topological phase diagram for ${\cal H}_F$ in the $\omega\rightarrow \infty$ limit. For the purpose of discussing Majorana physics in this work, the BdG Chern number $C$ for ${\cal H}_F$ [e.g. the ones shown in Fig. \ref{Fig2} (a) and (b)] is always defined for all quasienergy bands lying between $(-\omega/2,0]$. We expect the quasienergy band set within $(0,\omega/2]$ to carry an opposite $C$. While the Floquet chiral TSC in the high-frequency limit is adiabatically equivalent to a static chiral TSC with the same BdG Chern number, it serves as a good starting point for introducing the anomalous Floquet chiral TSC physics via quantum dynamics, which we do next. 

\begin{figure*}[t]
	\includegraphics[width=0.95\textwidth]{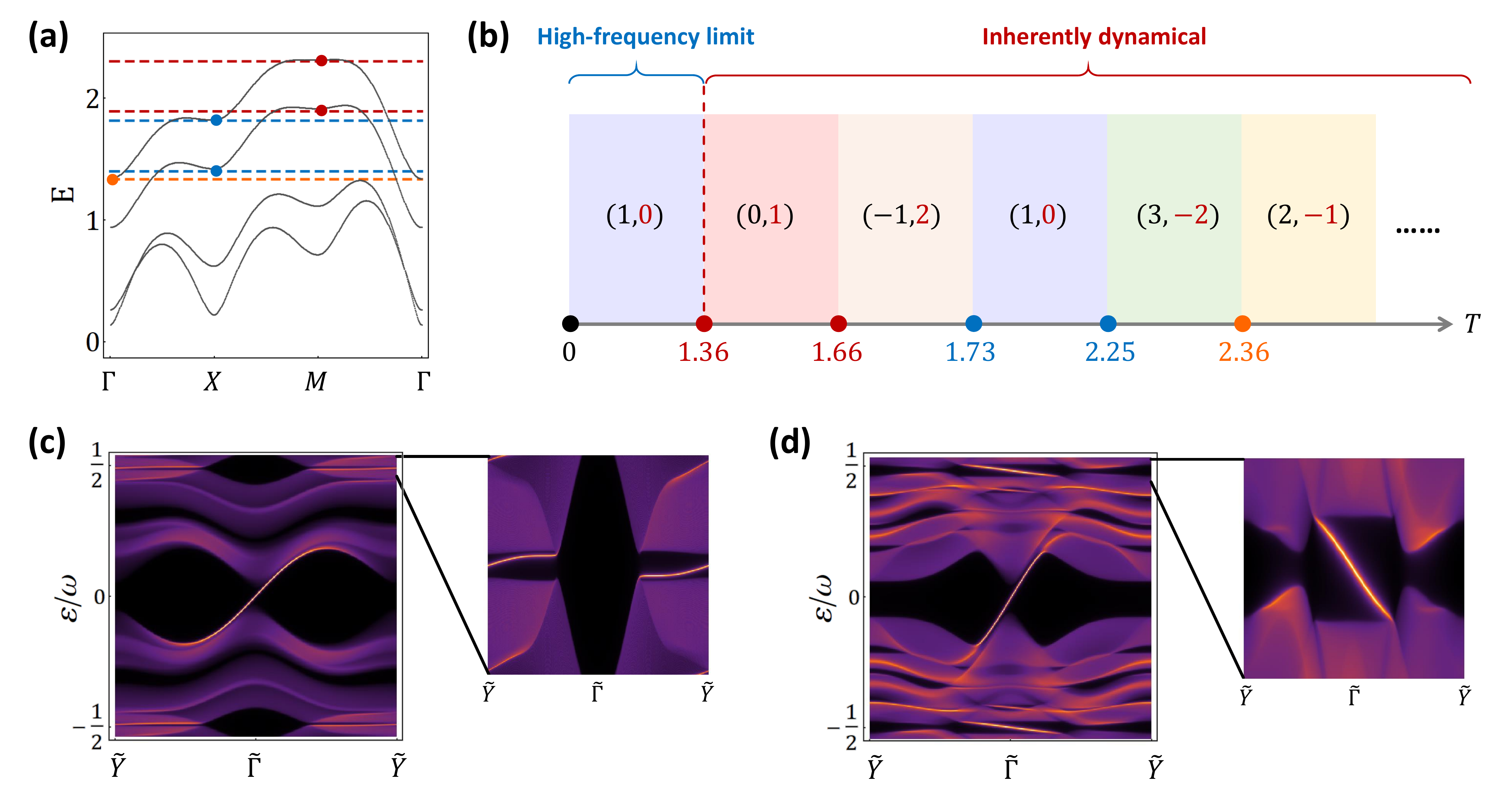}
	\caption{(a) The dynamical topological phase transition happens whenever the reference line of $\varepsilon=\omega/2$ intersects with the static bands of $h_\omega^{(0)}$ at high-symmetry momenta. Here we only plot the upper half spectrum of $h_\omega^{(0)}$ [i.e. the $E>0$ part of Fig. \ref{Fig2} (c)] for simplicity. (b) Driving-induced topological phase diagram as a function of the driving period $T$. Each phase is labeled by two integer-valued topological invariants: the BdG Chern number $C$ and the winding number ${\cal W}$. (c) Edge spectrum of the AFCTSC phase at $T=1.5$ with $(C, {\cal W})=(0,1)$. This phase features a single right-moving chiral Majorana edge mode penetrating the quasienergy gaps at both $\varepsilon=0$ and $\varepsilon=\omega/2$. (d) Another Floquet TSC phase with $(C, {\cal W})=(2,-1)$ occurs at $T=2.5$. This phase features a right-moving (left-moving) chiral Majorana edge mode within the $\varepsilon=0$ ($\varepsilon=\omega/2$) gap. The zoomed plots in (c) and (d) range from $0.45\omega$ to $0.55\omega$.}
	\label{Fig3}
\end{figure*} 

{\it Dynamical Topological Phase Transitions} - With the Floquet chiral TSC phase, we are now ready to move away from its high-frequency limit by gradually reducing the driving frequency $\omega$. Practically, this process simply amounts to reducing the bias voltage $V_0$ in the experimental setting. We note that the value of $\omega/2$ sets the boundary of the Floquet Brillouin zone (BZ) for the quasienergy $\varepsilon$. Therefore, whenever the reference line of $\varepsilon=\omega/2$ intersects with the energy bands of $h_\omega^{(0)}$ at a high-symmetry momentum $k_i$, the bulk Floquet gap at $k_i$ must close at $\varepsilon=\omega/2$ as well, inducing a {\it dynamical} topological phase transition (DTPT). Throughout our work, we will refer DTPT as the phase transition occurring on the Floquet BZ boundary $\varepsilon=\omega/2$.   

Without loss of generality, we start with the same Floquet chiral TSC phase in Fig. \ref{Fig2} (c) and (d) as an example to demonstrate the key physics here. By gradually increasing the driving period from $T=0$ to $T=2.5$ (in unit of $1/t_0$), five DTPTs take place successively as shown in Fig. \ref{Fig3} (a). In particular, we find that the first DTPT happening at $M$ has two seemingly ``contradictory" consequences: (i) it trivializes the original BdG Chern number $C$ for the bulk Floquet bands; (ii) it {\it cannot} trivialize the preexisting chiral Majorana edge mode around $\varepsilon=0$. Therefore, the only possible scenario compatible with the above contradictions is that the chiral Majorana edge mode must cross both the quasienergy gaps at $\varepsilon=0$ and $\varepsilon=\omega/2$ simultaneously, regardless of a vanishing $C$.

We then proceed to calculate the $\hat{y}$-directional Floquet edge spectrum at $T=1.5$, right after the first DTPT. As shown in Fig. \ref{Fig3} (c), we find a pair of chiral Majorana edge modes living inside both quasienergy gaps as expected, which confirms our conjectured edge scenario. Such a pair of anomalous chiral Majorana edge modes are exactly the defining characteristics of the {\it anomalous Floquet chiral TSC} phase that we have defined in the introduction. This AFCTSC phase should be clearly distinguished from the conventional Floquet chiral TSCs (like the one we find in the high-frequency limit), since its edge Majorana physics cannot be accounted for by calculating the BdG Chern number $C$. In fact, the relevant topological invariant for such AFCTSC is a homotopy-based winding number \cite{rudner2013anomalous,yao2017topo}
\begin{equation}
	{\cal W} = \frac{1}{8\pi^2}\int dt dk^2 \text{tr}[\epsilon^{l_1 l_2 l_3}\prod_{j=1}^{3}(\widetilde{U}^{-1}\partial_{l_j} \widetilde{U})] \in \mathbb{Z},
\end{equation} 
where $\epsilon^{l_1l_2l_3}$ is the Levi-Civita symbol for $l_{1,2,3}\in\{k_x,k_y,t\}$. Here $\widetilde{U}({\bf k}, t)$ is the micromotion operator for the driven system. Starting from the time-evolution unitary $U({\bf k}, t) = {\cal T}\text{exp}[-i\int_0^t H_\text{CTJJ}({\bf k}, t) dt]$ with ${\cal T}$ denoting the time ordering, we have $\widetilde{U}({\bf k}, t) = U({\bf k}, t) \times U({\bf k}, T)^{-t/T}$, where the logarithmic branch cut is chosen to be $\omega/2$ for our purpose. Crucially, it is both the BdG Chern number $C$ and the winding invariant ${\cal W}$ that together determine the complete topological properties of a general 2D class D Floquet TSC, including both the anomalous and non-anomalous ones, leading to a $\mathbb{Z}\times \mathbb{Z}$ topological classification.

The general relation between the bulk topological indices $(C,{\cal W})$ and the edge Majorana physics can be understood as follows. Let us first denote the numbers of chiral Majorana modes within the $\varepsilon=\gamma$ gap as $n_\text{edge}(\gamma)$ for $\gamma=0,\omega/2$. In particular, when $n_\text{edge}(\gamma)$ is positive (negative), it indicates $|n_\text{edge}|$ number of right-moving (left-moving) chiral Majorana modes. Then the bulk-edge correspondence is given by
\begin{eqnarray}
	n_\text{edge}(0) = C+{\cal W},\ \ n_\text{edge}(\frac{\omega}{2}) = {\cal W}.
	\label{eq:bulk-edge relation}
\end{eqnarray} 
Specifically, the anomalous chiral Majorana configuration in Fig. \ref{Fig3} (c) corresponds to $(C, {\cal W})=(0,1)$, which agrees with our calculations of  bulk topological indices.

One can similarly identify the topological nature of all Floquet phases induced by the other four DTPTs in Fig. \ref{Fig3} (a), by calculating both their topological indices and the corresponding edge quasienergy spectra. In general, we find that (i) a DTPT at $\Gamma$ or $M$ will change both $C$ and ${\cal W}$ by $\pm1$ simultaneously; (ii) a DTPT at both $X$ and $Y$ will change both $C$ and ${\cal W}$ by $\pm 2$. As a result, the driving induced topological phase diagram is mapped out in Fig. \ref{Fig3} (b), with each phase labeled by its topological indices $(C, {\cal W})$. As another example, we also calculate the edge spectrum of the $(C, {\cal W})=(2,-1)$ phase at $T=2.5$ in Fig. \ref{Fig3} (d), where a right-moving (left-moving) chiral Majorana edge mode is found with the $\varepsilon=0$ ($\varepsilon=\omega/2$) gap, in agreement with the prediction in Eq. \ref{eq:bulk-edge relation}.  

In general, the possible outcome of driving-induced DTPTs sensitively depends on the choice of static parameters (e.g. $g_z$, $\Delta_b$, $\alpha$, etc.), so that it is not possible to map out a complete topological phase diagram in the high-dimensional generic parameter space. Nevertheless, based on Eq. \ref{eq:bulk-edge relation}, we expect that by gradually reducing $V_0$ from the high-frequency limit, {\it the DTPT will always generate a Floquet state with chiral Majorana edge modes across the $\omega/2$ quasienergy gap}, which, crucially, is independent of (i) the topological nature of the initial high-frequency limit; (ii) the precise position of DTPT occuring in the momentum space. Namely, even starting from a topologically trivial state with $C=0$ [like the ones shown in Fig. \ref{Fig2} (a) and (b)], the Floquet engineering from Josephson phase modulation can always lead to various anomalous Floquet topological phenomena in CTJJ through simply changing the bias voltage $V_0$. 
The Floquet TSC physics with Majorana edge modes is generated entirely by the intrinsic dynamics arising from the DC-biased Josephson effect in the CTJJ.

{\it Conclusion} - To summarize, we have proposed a topological Josephson junction as a new experimentally feasible paradigm for achieving anomalous Floquet chiral Majorana modes in solid-state systems. The remarkable electrical tunability of CTJJ allows for the realization of Floquet chiral TSC physics, with the Josephson DC voltage being the key control knob. We note that the topological insulator-superconductor sandwich structures have already been fabricated experimentally with either Pb \cite{qu2012strong} or Nb \cite{zhang2018two} as the superconductor layers. Meanwhile, magnetic topological insulator-superconductor heterostructures have been realized in experiments with thin films of Cr-doped (Bi,Sb)$_2$Te$_3$ and Nb \cite{he2017chiral,shen2020spectroscopic,Kayyalha2020absence}. Therefore, we believe that the proposed CTJJ structure, as well as its Floquet chiral TSC physics, should soon be experimentally realizable. Finally, we mention that in the presence of an external magnetic field, our system should host vortex Majorana bound states at quasienergies $0$ and/or $\omega/2$, because of its chiral TSC nature. How to design new braiding protocols for these Floquet Majorana bound states to implement fault-tolerant logic gate operations is an intriguing challenge for the future.   

{\it Acknowledgment} - R.-X. Z thanks Zhi-Cheng Yang and Jiabin Yu for helpful discussions. This work is supported by the Laboratory of Physical Sciences. R.-X. Z acknowledges a JQI Postdoctoral Fellowship. 

{\it Note Added} - We became aware of a very recent work \cite{peng2020floquet} proposing a planar Josephson junction on a 2DEG system in order to achieve Floquet topological superconductivity in one dimension.

\bibliographystyle{apsrev4-1}
\bibliography{FloquetTSC}

\end{document}